\documentclass[12pt]{iopart}

\usepackage{graphicx}
\usepackage{cite}
\usepackage{xcolor}

\newcommand{\bd}[1]{ \mbox{\boldmath $#1$}}

\begin{document}

\title[]{Spin Hamiltonian with large fourth order terms: Triple well potentials and Bloch sphere visualization}

\author{D S Lohr-Robles$^{1}$, M Grether$^{2}$, E L\'opez-Moreno$^{2}$, P O Hess$^{3,4}$}
\address{$^{1}$Independent researcher, Mexico}

\address{$^{2}$Facultad de Ciencias,  Universidad Nacional Aut\'onoma de M\'exico, 04510 Mexico-City, Mexico}

\address{$^{3}$Instituto de Ciencias Nucleares, Universidad Nacional Aut\'onoma de M\'exico, A.P. 70-543, 04510 Mexico-City, Mexico}

\address{$^{4}$Frankfurt Institute for Advanced Studies, J. W. von Goethe University, Hessen, Germany}
\ead{hess@nucleares.unam.mx}

\begin{abstract}
We present a study of a general spin Hamiltonian with terms up to fourth order. With the coherent states the semiclassical potential is obtained and with catastrophe theory its parameter space is constructed. When the fourth order parameters are large enough the parameter space has regions where the semiclassical potential has three wells. By applying an oscillating magnetic field a trajectory in parameter space crosses the Maxwell set multiple times resulting in many ground state quantum phase transitions. Using the coherent states we are able to visualize the localization of the ground state on the Bloch sphere as the magnetic field is varied.
\end{abstract}

{\small Conflict of Interest declaration: The authors declare that they have NO
affiliations with or involvement in any organization or entity with any
financial interest in the subject matter or materials discussed in this
manuscript.}

%
\vspace{2pc}

\section{Introduction}\label{introduction}
The use of spin Hamiltonian to model many body systems has experienced a wide variety of applications in different areas of physics. By an appropriate fitting of the interacting parameters the phenomena of interest may be accurately described, assigning physical meaning to each of the terms. For this reason, theoretical work when the Hamiltonian has higher order contributing terms can be of use, in particular when the value of the corresponding parameters are large enough to induce a meaningful change, and how this contribution affects the phase transitions of the system. The physics of single molecule magnets (SMMs) can be accurately described using a spin Hamiltonian \cite{nanomagnets}. The applications of SMMs in new technologies as qubits remains a possibility \cite{jiang2012a,jenkins2017a,morenopineda2018a,gaitaarino2019a,morenopineda2021a,zhang2021a}.

Spin Hamiltonians have also been used in the description of quantum phase transitions (QPTs) and excited states quantum phase transitions (ESQPTs) as a function of the Hamiltonian parameters \cite{cejnar2021a}, 
and they have also been applied to model systems of very large spin in ferritin proteins \cite{hagen2024a}.
As a result these models are very useful as a way to test new phenomena. The motivation of the present contribution is to study phase transitions in SMMs and the search of a tractable manner to manipulate them.

The present paper is structured as follows: In section \ref{section2} the spin Hamiltonian is introduced with terms up to fourth order, the coherent states are defined and the parameter space of the semiclassical potential is obtained. In section \ref{section3} the master equation that defines the population of states as a function of time is introduced, together with the concept of relaxation time of the material. In section \ref{section4} the concept of fidelity and fidelity susceptibility and their relation to QPTs is discussed. A series of results for three different examples are presented in section \ref{section5}. In section \ref{section6}, hypothetical examples of systems with large values of the fourth order parameters are considered, which lead to systems with three stability wells in the semiclassical potentials; two of these examples are presented and studied. The concept of the Bloch sphere is introduced as a way to visualize and describe the localization of the eigenstates of the Hamiltonian with the intention to {\it construct and manipulate qubits}. Finally, in section \ref{conclusions} conclusions are drawn and possible future work is discussed.

\section{Spin Hamiltonian}\label{section2}
We consider the following expression for a spin 
Hamiltonian, designed to describe magnetic molecules, with operators up to fourth order:
\begin{eqnarray}\label{1.1}
{\bd H} &=& - \frac{g \mu_B}{S} \vec{{\bd B}}\cdot\vec{{\bd S}} + \frac{1}{S(2S-1)} \left( D \left({\bd S}^2_z - \frac{1}{3} S(S+1)\right) + E({\bd S}^2_{x}-{\bd S}^2_{y}) \right) \cr
&& + \frac{{\bd H}_4}{S(2S-1)(2S-2)(2S-3)} 
\end{eqnarray}
where $g$ is the Land\'e factor, $\mu_B$ is the Bohr magneton, $D$ and $E$ are the second order anisotropy constants of the molecule and ${\bd H}_4$ indicates fourth order anisotropy terms. The first order interaction is given by the Zeeman term in (\ref{1.1}), with $\vec{{\bd B}}=(B_x,B_y,B_z)$ an external magnetic field and $\vec{{\bd S}}=(S_x,S_y,S_z)$ the spin vector, with $S_{i}$, $i=x,y,z$, the spin components operators. The explicit form of the fourth order terms is given by:
\begin{eqnarray}\label{1.1a}
\fl {\bd H}_4 &=& B_4^0 {\bd O}_4^0 + B_4^2 {\bd O}_4^2 + B_4^3 {\bd O}_4^3 + B_4^4 {\bd O}_4^4 \cr
\fl &=& B_4^0 \left(35{\bd S}_z^4 + (25-30S(S+1)){\bd S}_z^2 +3S^2(S+1)^2 -6 S(S+1) \right) \cr
\fl && +\frac{B_4^2}{4} \left((7{\bd S}_z^2 - S(S+1)-5)({\bd S}_{+}^2 + {\bd S}_{-}^2)  +({\bd S}_{+}^2 + {\bd S}_{-}^2)(7{\bd S}_z^2 - S(S+1)-5) \right) \cr
\fl &&   +\frac{B_4^3}{4} \left( {\bd S}_z ({\bd S}_{+}^3 + {\bd S}_{-}^3)+({\bd S}_{+}^3 + {\bd S}_{-}^3){\bd S}_z \right) + \frac{B_4^4}{2}({\bd S}_{+}^4 + {\bd S}_{-}^4),
\end{eqnarray}
where ${\bd O}_q^k$ are the Steven operators \cite{nanomagnets,gatteschi2003a,cornia2001a} and $B_q^k$ the corresponding fourth order anisotropy constants. It is important to notice that the values of these constants are smaller than that of the second order ones, and the non-zero values are dependent on the symmetric properties of the material: tetragonal symmetry ($k=0,4$), orthorhombic symmetry ($k=0,2,4$), and trigonal symmetry ($k=0,3$) \cite{gatteschi2003a,cornia2001a}. To each of the terms in the Hamiltonian we add a factor of the form $2\prod_q(2S-q+1)^{-1}$ for the $q$-th order interaction term. This is to ensure that the semiclassical potential calculated in the next subsection is independent of $S$.

Using the complete basis $|SM\rangle$ of the eigenvectors of the $S_z$ operator, satisfying the eigenvalue equation $S_z |SM\rangle = M\hbar |SM\rangle$, the Hamiltonian matrix of (\ref{1.1}) is constructed and its eigenvalues $E_k$, which satisfy
\begin{equation}\label{1.2}
{\bd H} |\psi_k\rangle = E_k |\psi_k\rangle ,
\end{equation}
are obtained by diagonalization. The eigenvectors $|\psi_k\rangle$ are expressed as linear combinations in the $|SM\rangle$ basis as:
\begin{equation}\label{1.3}
|\psi_k\rangle = \sum_{M=-S}^{S} c_{k,M}|SM\rangle,
\end{equation}
where the square of the absolute value of the coefficients $c_{k,M}$ represent the probability of finding the state $|\psi_k\rangle$ with spin projection $M$, with $M=-S,-S+1,\ldots,S-1,S$. In the following we will simplify the notation defining $|SM\rangle \equiv |M\rangle$.

The Hamiltonian of the system written in (\ref{1.1}) depends on nine parameters: Three free parameters for the interaction with the magnetic field and six anisotropic constants (two second order and four fourth order). The non-free parameters are adjusted to experiment. We want to study the effect of these parameters in the structure of the eigenstates of the Hamiltonian. Using the $\mathrm{SU}(2)$ coherent states \cite{arecchi1972a} we can obtain the semiclassical potential of the system as the expectation value of the Hamiltonian in the coherent state basis. We will find that by studying the critical points of the semiclassical potential we can obtain information about the behaviour of the eigenvalues and eigenvectors. Catastrophe theory is a very useful method 
\cite{thom,gilmore,arnold} for the categorization of functions that depend on parameters, as it presents a classification of the stable singularities of a potential functions leading to the construction of separatrices in parameter space, denoting the different phases of the system, identifiable as elementary catastrophes: fold, cusp, swallowtail, butterfly, etc., and their respective regions of stability. The catastrophe theory has been applied with great success to nuclear and particle physics \cite{lopezmoreno1996a,lohrrobles2021a}.

\subsection{Coherent states and semiclassical potential}\label{subsection2.1}
The semiclassical potential is obtained by calculating the expectation value of the Hamiltonian in the atomic coherent state basis: $V(\theta,\phi) = \langle \zeta | {\bd H} |\zeta \rangle$, where the coherent state $|\zeta\rangle$ is defined as a rotation of lowest weighted state $|-S\rangle$ of angle $\theta$ about an axis $\hat{n}=(\sin\phi,-\cos\phi,0)$ in angular momentum space \cite{arecchi1972a}:
\begin{eqnarray}\label{2.1}
\fl |\zeta \rangle &=& R(\theta,\phi) |-S\rangle \cr
\fl &=& (1+ |\zeta|^2)^{-S}e^{\zeta S_{+}} |-S\rangle  \cr
\fl &=& \sum_{M=-S}^S \left(\frac{(2S)!}{(S+M)!(S-M)!}\right)^{1/2}\left(\cos \frac{\theta}{2}\right)^{S-M}\left(e^{-\mathrm{i} \phi}\sin \frac{\theta}{2}\right)^{S+M} |M\rangle
\end{eqnarray}
with $\zeta = e^{-i\phi}\tan(\theta/2)$.

The semiclassical potential is explicitly given by:
\begin{eqnarray}\label{2.4}
\fl  V(\theta,\phi;D,E,B_i) &=& \frac{D}{12}(1+3\cos 2\theta) + \frac{E}{2}\cos 2\phi\sin^2 \theta \cr
\fl & & - g\mu_B \left(-B_z \cos \theta + B_x\cos\phi \sin\theta + B_y \sin\phi \sin\theta \right) \cr
\fl & & +\frac{1}{8}\left(\frac{B_4^0}{8}(35\cos 4\theta  +20\cos 2\theta +9)  + \frac{B_4^2}{2}(7\cos 2\theta +5)\cos 2\phi \sin^2\theta  \right. \cr
\fl & &\left.   - B_4^3 \cos 3\phi \cos\theta \sin^3\theta + B_4^4 \cos 4\phi \sin^4\theta \right) ,
\end{eqnarray}
which is a function of two angular variables $(\theta,\phi)$ and independent of the value of spin $S$ of the system. Therefore, the parameter spaces obtained in these sections are {\it valid for all values of $S$}. Because our interest lies on how the changes of parameters can lead the system to change from one phase to another, we will find that very telling results, e.g. indications of QPTs, can be obtained even for small values of $S$.

As a starting point we will restrict ourselves to the case when the magnetic field is constrained in the $xz$-plane: $\vec{B}=(B_x,0,B_z)$, as previously done in \cite{lohrrobles2023b}. We can see in (\ref{2.4}) that when $B_y=0$, the values $\phi_c=0,\pi$, are critical points for all values of the parameters. Thus, we are able to substitute these values in (\ref{2.4}) and focus on the study of the following one-dimensional potential function:
\begin{eqnarray}\label{2.5}
 V(\theta,\phi_c;r_i) &=& r_1 \cos\phi_c\sin\theta + r_2 \cos \theta  + r_3 \cos 2\theta  +  r_4 \cos 4\theta \cr
&&+ r_5 \cos\phi_c (2\sin 2\theta - \sin 4\theta), 
\end{eqnarray}
where we defined the new parameters $r_i$ as:
\begin{eqnarray}\label{2.6}
r_1 &=& -g\mu_BB_x \cr
r_2 &=& g\mu_BB_z \cr
r_3 &=& \frac{1}{4} (D-E)+\frac{1}{16}(5B_4^0+B_4^2-B_4^4) \cr
r_4 &=& \frac{1}{64}(35 B_4^0 -7B_4^2 + B_4^4 )\cr
r_5 &=&-\frac{1}{64} B_4^3 .
\end{eqnarray}
Inversely, once the parameters $r_i$ are known, the magnetic field components and $D$, $E$ can be
deduced.

\subsection{Bifurcation and Maxwell sets}\label{subsection2.2}
The bifurcation set is a subspace in parameter space where critical points of the potential function begin to emerge. Thus, in its vicinity we can find two phases (regions), in one there exist a stability point $\theta_c$, while in the other there is no such point. The bifurcation set can be found considering the critical manifold, which is the hypersurface of critical points $\theta_c$ satisfying:
\begin{equation}\label{2.1.1}
\left.\frac{d}{d \theta} V(\theta,\phi_c;r_i)\right|_{\theta=\theta_c} =0
\end{equation}
spanned by a continuous variation of the parameters $r_i$. The set of points where the mapping of the critical manifold to the parameter space is singular is defined as the bifurcation set. A straightforward calculation of this singular mapping result in the following set of parametric functions:
\begin{eqnarray}\label{2.1.2}
\fl  r_1 (x,\phi_c;r_3,r_4,r_5)&=& r_3 \cos\phi_c (3\sin x - \sin 3x) + 2r_4 \cos\phi_c (5\sin 3x - 3\sin 5x)\cr
\fl && - 6 r_5 (\cos x -2\cos 3x + \cos 5x)\cr
\fl  r_2 (x,\phi_c;r_3,r_4,r_5)&=& -r_3 (3\cos x +\cos 3x)-2r_4 (5\cos 3x + 3\cos 5x) \cr
\fl &&+ 4 r_5 \cos \phi_c \sin x(2+7\cos 2x +3 \cos 4x),
\end{eqnarray}
where $x=\theta_c$ is the critical point and satisfies (\ref{2.1.1}). These parametric functions allow us to draw the bifurcation set in $(r_1,r_2)$ parameter space for a given set of values $(r_3,r_4,r_5)$. One can also solve these equations to obtain the bifurcation set in any other $(r_i,r_j)$ parameter space of interest.

The Maxwell set is the subspace in parameter space where two or more extrema (minima or maxima) of the potential function have the same value, i.e. $V(\theta_1,\phi_c;r_i)=V(\theta_2,\phi_c;r_i)$; thus in its vicinity two phases exist, in one $V(\theta_1)>V(\theta_2)$, while in the other $V(\theta_1)<V(\theta_2)$, i.e. there is a change of the dominant stability point (phase). Similarly to the bifurcation set, the Maxwell set can be found by considering the singular mapping of a hypersurface to the parameter space. The particular hypersurface is defined as the set of roots $\theta_i$, such that $V(\theta_i,\phi_c; r_i)+V_0=0$, where $V_0$ is an arbitrary real number, spanned by a continuous variation of the parameters $r_i$. Then one has to find the values of $r_i$ such that the above description is true for two values $\theta_1$ and $\theta_2$. Performing the calculation is very involved. Fortunately, using the routines of the software MATHEMATICA we could resolve the problem and we found the following set of functions for the Maxwell set:
\begin{eqnarray}\label{2.1.3}
\fl  r_1 (\theta_1,\theta_2;r_4,r_5)&= 8\sec\phi_c \sin \theta_1 \sin \theta_2 (\sin \theta_1 +\sin \theta_2) \Bigg(2r_4 (1-\cos(\theta_1 + \theta_2)) \cr
 &\quad +r_5 \cos 3\phi_c \bigg(2\cot(\theta_1 + \theta_2)-\cos 2(\theta_1 + \theta_2)\csc(\theta_1 + \theta_2) \bigg) \Bigg) \cr
\fl r_2 (\theta_1,\theta_2;r_4,r_5)&= 2 \cos\frac{\theta_1 - \theta_2}{2}\Bigg(32 r_4 \cos \theta_1 \cos \theta_2 \cos^3 \frac{\theta_1 + \theta_2}{2} \Bigg)\cr
 &\quad - r_5 \cos 3\phi_c \csc \frac{\theta_1 + \theta_2}{2} \bigg(2 -2 \cos 2\theta_1 - 2\cos 2\theta_2 - \cos(\theta_1 + \theta_2) \cr
 &\quad - 2 \cos 2(\theta_1 + \theta_2) - \cos 3(\theta_1 + \theta_2) - \cos(3\theta_1 + \theta_2)-\cos(\theta_1 +3\theta_2) \bigg) \cr
\fl r_3 (\theta_1,\theta_2;r_4,r_5)&= -2 r_4 \bigg(3 \cos 2\theta_1 +3\cos 2\theta_2 + 4\cos(\theta_1 + \theta_2) \bigg)\cr
&\quad + \frac{r_5}{2}\cos 3\phi_c \csc \frac{\theta_1 + \theta_2}{2} \sec \frac{\theta_1 + \theta_2}{2} \bigg(2\cos(\theta_1 + \theta_2)  -4\cos(2\theta_1 +2\theta_2)\cr
&\quad -3 \big(\cos(3\theta_1 +\theta_2)+\cos(\theta_1 +3\theta_2)\big) \bigg)
\end{eqnarray}
and we are able to draw the Maxwell set as a parametric function in parameter space $(r_1,r_2)$ given the critical points $(\theta_1,\theta_2)$ that satisfy (\ref{2.1.3}) for a given set of values $(r_3,r_4,r_5)$.

In the following applications, examples of phases are shown, and the corresponding separatrices of the bifurcation and Maxwell sets will be illustrated.

\section{Magnetization and master equation}\label{section3}
The labelling of the eigenstates of the Hamiltonian (\ref{1.1}) is an important issue to address. In cases when the $D$ parameter is the most relevant the low energy states have a dominant coefficient $c_{k,M}$ in the expression (\ref{1.3}), while the other coefficients are small. In this cases it is possible to maintain the labelling of states of the spin projections, which is required when studying transitions between states in the interaction of the system with a magnetic field. In this framework the states labelled $M=\pm S$, are the lowest states in their respective potential wells.

The population $p_m$ of the state labelled $m$ at a time $t$ is described by the master equation \cite{nanomagnets}:
\begin{equation}\label{3.1}
\frac{d p_m(t)}{d t} = \sum_{m'} \big(\gamma_{m'm} p_{m'}(t) - \gamma_{mm'}p_{m}(t) \big)
\end{equation}
with $\gamma_{mm'}$ the transition rates of going from the state $m'$ to $m$, and are given by 
\begin{eqnarray}\label{3.2}
\fl \gamma_{mm'} &=& \frac{3}{\pi \hbar^4 \rho c_s^5}\frac{(E_{m'} - E_{m})^3}{e^{(E_{m'}-E_{m})/k_B T}-1} \Big( |D_1|^2 \big(|\langle \psi_{m'}| S_+^2 |\psi_{m}\rangle|^2 +|\langle \psi_{m'}| S_-^2 |\psi_{m}\rangle|^2 \big) \cr
\fl &&  + |D_2|^2 \big(|\langle \psi_{m'}|\{S_+,S_z\} |\psi_{m}\rangle|^2 +|\langle \psi_{m'}|\{S_-,S_z\} |\psi_{m}\rangle|^2 \big)   \Big)
\end{eqnarray}
where $\rho$ is the density of the material, $c_s$ is the velocity of sound in the material, $D_1$ and $D_2$ the spin-phonon couplings parameters, and $\{\cdot,\cdot\}$ is the anti-commutator \cite{nanomagnets,mannini2010a}.

The differential equation in (\ref{3.1}) can be written in matrix form as:
\begin{equation}\label{3.3}
\frac{d \vec{p}(t)}{d t} =  \vec{p}(t) {\bd G}
\end{equation}
where ${\bd G}$ is the transition rate matrix, and its respective matrix elements are given by:
\begin{equation}\label{3.4}
G_{mm'} =\gamma_{mm'}-\delta_{mm'}\sum_{k}\gamma_{mk} .
\end{equation}

In order to solve the master equation (\ref{3.1}) one has to take into account that the variation of the magnetic field in (\ref{1.1}) is time dependent: $|\vec{B}|=\nu t + B_0$, with $\nu = dB/dt$ the sweeping rate of magnetization and $B_0$ the initial magnetization. 
The study of the master equation has been considered for various SMMs in order to model the magnetization hysteresis loops \cite{fernandez1998a,luis1998a,mannini2010a,serrano2020a}. Monte Carlo methods have proven to be a strong computational technique when dealing with time dependent transition rates \cite{fichthorn1991a,jansen1995a,liu2009a,liu2010a,cuppen2013a}. In particular, in \cite{serrano2020a} they used a kinetic Montecarlo variable size step method to solve the master equation, which we will follow in the present paper.

The magnetization hysteresis loops are obtained using the solutions of the master equation (\ref{3.1}) as: 
\begin{equation}\label{3.5}
M = \sum_{m} p_m(t) \langle \psi_m |S_z| \psi_m \rangle .
\end{equation}

\subsection{Relaxation time and relaxation rate}\label{section3.1}
Measurements of the relaxation time and rate as a function of the magnetic field have been of considerable importance in the understanding of SMMs \cite{thomas1996a,wernsdorfer2000a,ueda2002a,dressel2002a,adams2013a}. 
The relaxation time $\tau$ of the material as a function of the magnetic field can be obtained with the eigenvalues of the transition rate matrix in (\ref{3.4}) as done in \cite{nanomagnets,villain1994a,leuenberger1999a,leuenberger2000a}. It is proven in \cite{nanomagnets} that one of the eigenvalues is equal to zero and corresponds to the equilibrium state, while the others are negative. The relaxation time is then identified as the reciprocal of the smallest non-zero eigenvalue $g_{min}$ \cite{leuenberger1999a}:
\begin{equation}\label{3.1.1}
\tau = -\frac{1}{g_{min}} .
\end{equation}
The relaxation time tells us how long does the magnetization due to the alignment of the molecules persists in the presence of the magnetic field.

The relaxation rate $\Gamma$ is in turn the reciprocal of the relaxation time \cite{nanomagnets,leuenberger1999a}:
\begin{equation}\label{3.1.2}
\Gamma = \frac{1}{\tau}
\end{equation}
In \cite{leuenberger1999a} the peaks of the relaxation rate as a function of time are fitted with a Lorentzian function.

\section{Fidelity and fidelity susceptibility}\label{section4}
Quantum fidelity is a measure of how much does a quantum state, that depends on a parameter, resembles itself with a small 
variation of the parameter \cite{gu2010a}, e.g. the magnetic field amplitude. This quantity has proven to be very useful for the identification of QPTs, characterized by a sudden drop in fidelity at the critical point of the parameter, and has been used successfully to detect QPTs in quantum many-body systems \cite{gu2010a,zanardi2006a,buonsante2007a,you2007a,tzeng2008a,tian2011a,plotz2011a,rams2011a}. In order to calculate the quantum fidelity of a state as a function of the parameter one needs to introduce an arbitrary small value $dB$.

The fidelity of a state $|\psi_k\rangle$ is defined as \cite{gu2010a}:
\begin{equation}\label{4.1}
F_{k}= |\langle \psi_k (B_i-dB_i)|\psi_k(B_i+dB_i)\rangle|^2
\end{equation}
where $B_i$ is the magnetic field parameter and $dB_i$ represents a small increment. Using this expression one is able to calculate the value of the fidelity of the $k$-th state as a function of a varying magnetic field.

The quantum fidelity susceptibility $\chi_{F_k}$ is defined as the second order coefficient of the Taylor series expansion of the quantum fidelity in (\ref{4.1}) about $dB=0$ \cite{gu2010a}. It contains all the information of the quantum fidelity and has the advantage that it is independent of the arbitrary value of $dB$, i.e. it does not depend on any external value and it is therefore more suitable for calculations. 

The fidelity susceptibility can be explicitly written as \cite{gu2010a}
\begin{equation}\label{4.2}
\chi_{F_k} = 2 \sum_{m\neq k} \frac{|\langle \psi_m |S_z|\psi_k\rangle|^2}{(E_m - E_k)^2}
\end{equation}
where $S_z$ is related to the parameter of the interaction, i.e. the magnetic field.

\section{Results}\label{section5}
In this section we will present results of the physical phenomena describe in the previous sections for three different test cases of magnetic molecules: 
a) $\mathrm{Fe}_8$ SMM, b) $\mathrm{Fe}_4$ SMM, c) Arbitrary parameters, not related to a specific magnetic molecule, with a large fourth order term.

\subsection{$\mathrm{Fe}_8$ SMM}
We start with the parameter values given in the review \cite{gatteschi2003a} and modify them a little in order to better fit the hysteresis plots in \cite{wernsdorfer2000b}: $S=10$,  $D/k_B=-0.295$ K, $E/k_B=0.056$ K, $B_4^0/k_B=1.15\times 10^{-6}$ K, $B_4^2 /k_B =-1.15\times 10^{-6}$ K, $B_4^4 /k_B =-2.18\times 10^{-5}$ K, $ B_x =0.02$ T, $B_y=B_4^3=0$. In this case we considered the magnetic field to be not completely aligned with the $z$-axis and the easy magnetization axis of the molecules, but making a small $3^{\circ}$ angle with respect to it; this was done to recreate the hysteresis steps in the presence of a very small transverse field.

The parameters in the transition rates $\gamma_{mm'}$ in (\ref{3.2}) used are: $3/(\pi \hbar^4 \rho c_s^5) = 3.13\times 10^3\;\mathrm{K}^{-5}\mathrm{s}^{-1}$, $D_1=D_2=0.26$ K, and $T=0.04$ K.

\begin{figure}[h!]
\begin{center}
\includegraphics[scale=0.55]{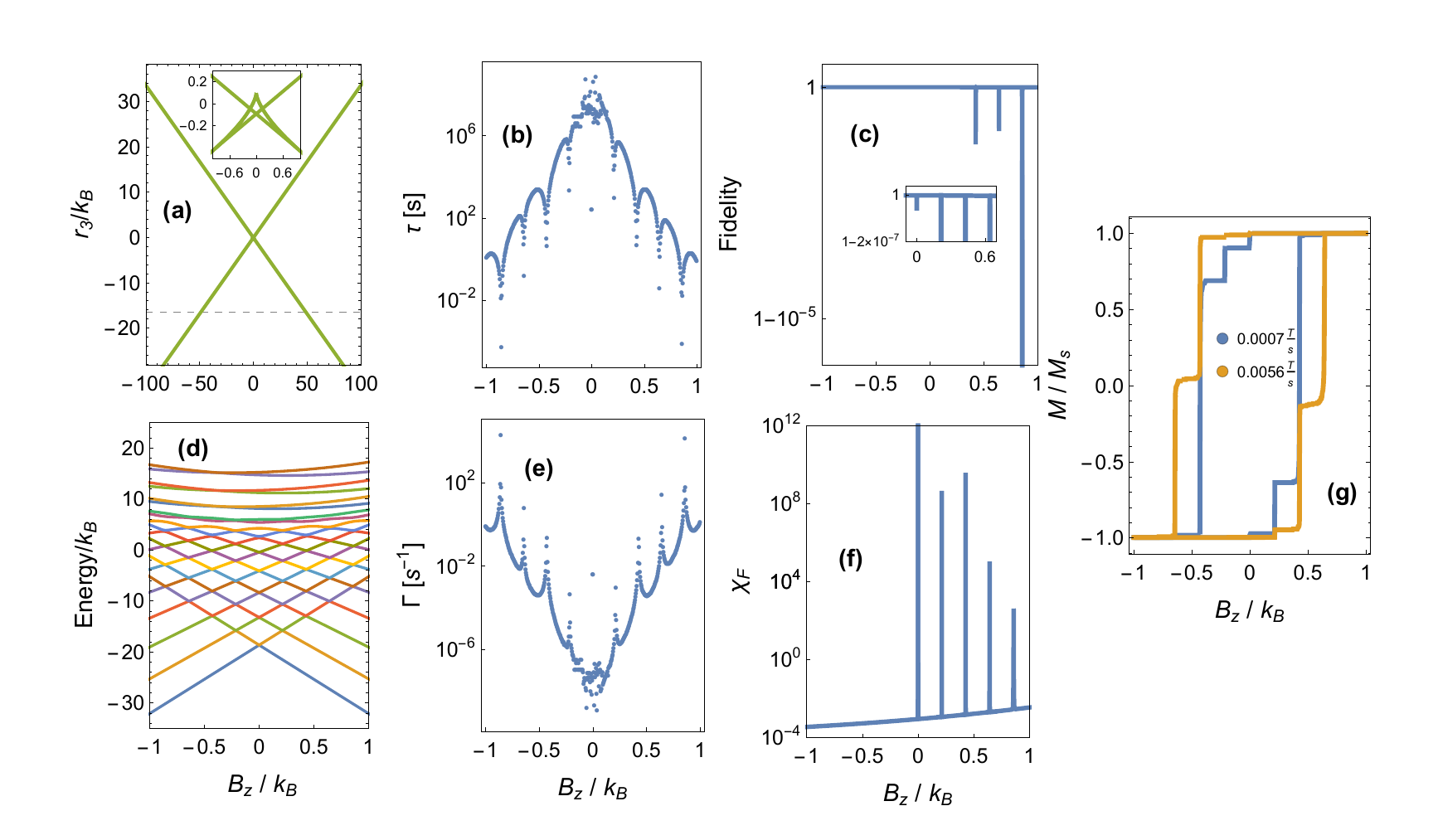}
\end{center}
\caption{(a) Bifurcation set in $(r_2/k_B,r_3/k_B)$ parameter space and the dotted line represents the parameters for the $\mathrm{Fe}_8$ example: $S=10$,  $D/k_B=-0.295$ K, $E/k_B=0.056$ K, $B_4^0/k_B=1.15\times 10^{-6}$ K, $B_4^2 /k_B =-1.15\times 10^{-6}$ K, $B_4^4 /k_B =-2.18\times 10^{-5}$ K, $ B_x =0.01$ T, $B_y=B_4^3=0$; and $3/(\pi \hbar^4 \rho c_s^5) = 3.13\times 10^3\;\mathrm{K}^{-5}\mathrm{s}^{-1}$, $D_1=D_2=0.26$ K, and $T=0.04$ K. (b) Relaxation time, (c) fidelity, (d) energy levels, (e) relaxation rate, (f) fidelity susceptibility, (g) hysteresis loop for sweeping rates: $0.007$ T/s and $0.056$ T/s, as a functions of $B_z$. To calculate the fidelity we used $dB_z=0.001$.}
\label{fig.1}
\end{figure}

In figure \ref{fig.1}(a) we show the bifurcation set, and the horizontal dashed line corresponds to the parameter values of $\mathrm{Fe}_8$. When the bifurcation set is crossed the avoided level crossings of higher energy levels start to occur, as a consequence of the double well structure of the semiclassical energy surface. At zero magnetic field both potential wells are at the same depth and there is an avoided level crossing of the ground state, which can be seen in figure \ref{fig.1}(d). In figures \ref{fig.1}(b) and \ref{fig.1}(e) we plotted the relaxation time and relaxation rate as functions of the magnetic field, respectively. The drops of the relaxation time, which are peaks in the relaxation rate, occur at the avoided level crossings of the eigenstate with label $m=-10$, as seen in figure \ref{fig.1}(d). A similar behaviour occur with the fidelity and fidelity susceptibility seen in figures \ref{fig.1}(c) and \ref{fig.1}(f), respectively. 
In figure \ref{fig.1}(g) we plotted the hysteresis loop for two values of the sweeping rate of the magnetic field in the $z$ direction. These results can be compared with the experimental results obtained in figure 2 in \cite{wernsdorfer2000b}. The transition rates defined in (\ref{3.2}) for the master equation in this model does not adequately reproduce the transitions at zero magnetic field for pure tunneling, i.e. at low temperature, as described in \cite{serrano2020a}, and to solve this issue and additional constant correction term is added to the transition rates $\gamma_{mm'}$ for these cases. Following the methodology described in \cite{serrano2020a}, we added the constant correction term $\gamma_t=0.002\;\mathrm{s^{-1}}$ in order to obtain the hysteresis in figure \ref{fig.1}(g). The location of the steps in the hysteresis loops also coincide with the location of the drops of the relaxation time.

\subsection{$\mathrm{Fe}_4$ SMM}
We use the parameters obtained in \cite{vergnani2012a}, with the addition of a fourth order term $B_4^3$ permitted by the tetragonal symmetry of the molecule to better fit the experimental hysteresis loops. The parameters used for $\mathrm{Fe}_4$ were: $S=5$,  $D/k_B=-0.601$ K, $E/k_B=0.024$ K, $B_4^0/k_B=2.88\times 10^{-5}$ K, $B_4^3 /k_B = 0.0004$ K, $ B_x =B_y=B_4^2=B_4^4=0$.

The parameters in the transition rates $\gamma_{mm'}$ in (\ref{3.2}) used are: $3/(\pi \hbar^4 \rho c_s^5) = 4.1\times 10^3\;\mathrm{K}^{-5}\mathrm{s}^{-1}$, $D_1=D_2=0.28$ K, and $T=0.04$ K. For this case we also add a constant correction term $\gamma_t$ to the transition rate to enhance the tunnelling between states $m=\pm 5$ and $m=\pm 4$, as was done in \cite{serrano2020a}. Depending on the magnetic sweeping rate we have: $\gamma_t = 0.1\; \mathrm{s}^{-1}$ for the sweeping rate $0.001\; \mathrm{T}\mathrm{s}^{-1}$ and $\gamma_t = 0.6\; \mathrm{s}^{-1}$ for the sweeping rate $0.017\; \mathrm{T}\mathrm{s}^{-1}$. 

\begin{figure}[h!]
\begin{center}
\includegraphics[scale=0.6]{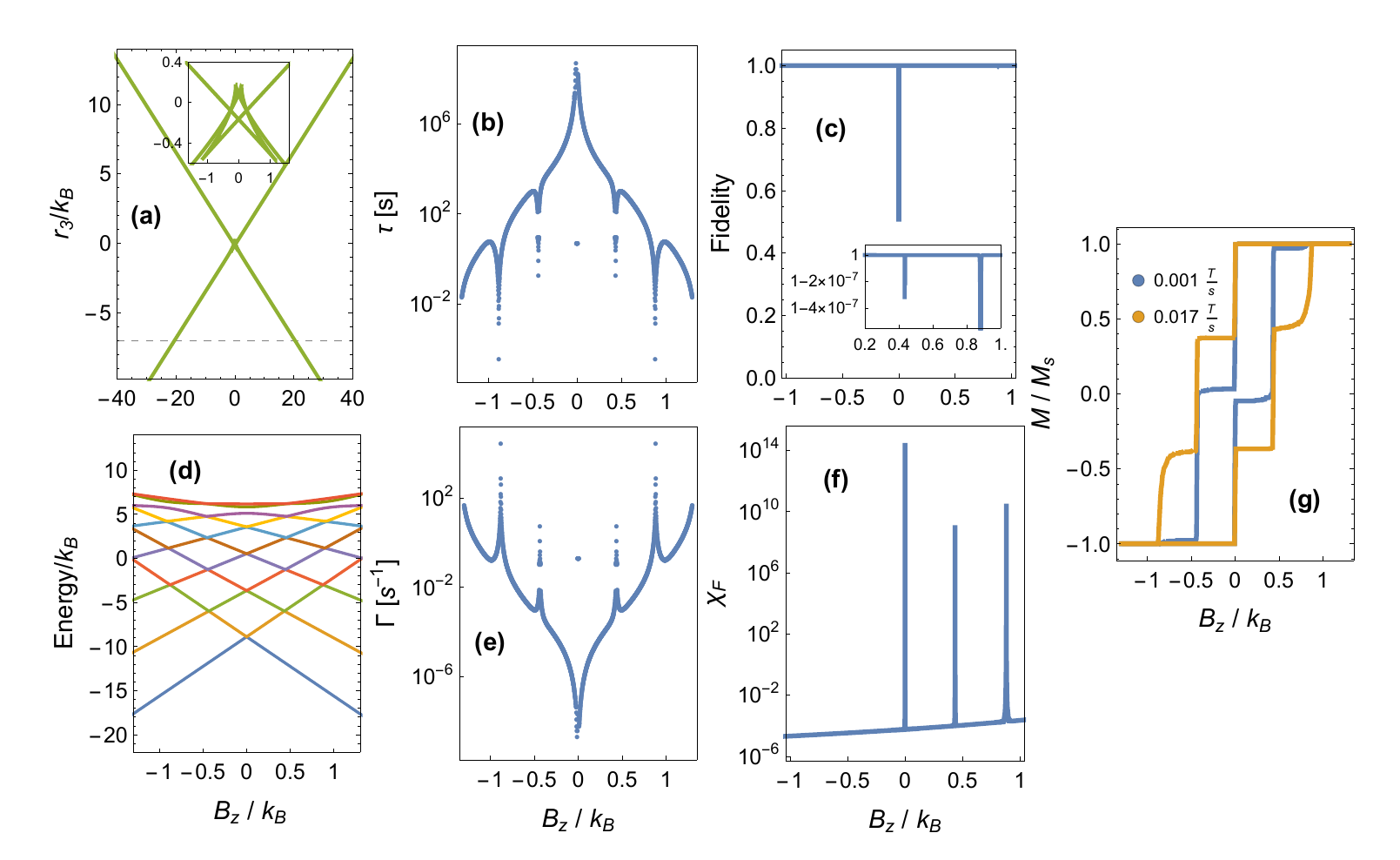}
\end{center}
\caption{(a) Bifurcation set in $(r_2/k_B,r_3/k_B)$ parameter space and the dotted line represents the parameters for the $\mathrm{Fe}_4$ example: $S=5$,  $D/k_B=-0.601$ K, $E/k_B=0.024$ K, $B_4^0/k_B=2.88\times 10^{-5}$ K, $B_4^3 /k_B = 0.0004$ K, $ B_x =B_y=B_4^2=B_4^4=0$; and $3/(\pi \hbar^4 \rho c_s^5) = 4.1\times 10^3\;\mathrm{K}^{-5}\mathrm{s}^{-1}$, $D_1=D_2=0.28$ K, and $T=0.04$ K. (b) Relaxation time, (c) fidelity, (d) energy levels, (e) relaxation rate, (f) fidelity susceptibility, (g) hysteresis loop for sweeping rates: $0.001$ T/s and $0.017$ T/s, as a functions of $B_z$. To calculate the fidelity we used $dB_z=0.001$.}
\label{fig.2}
\end{figure}

In figure \ref{fig.2} we show the same results as in figure \ref{fig.1} but for the $\mathrm{Fe}_4$ SMM. In this case we can see in figures \ref{fig.2}(b) and \ref{fig.2}(e) the consequences of the constant correction term for the transition rate between states $m=-5$ and $m=5$ at zero magnetic field, were the value of the relaxation time is very small in the vicinity of $B_z=0$, while the relaxation rate has an artificial peak in that vicinity. 
In figure \ref{fig.2}(g) we plotted the hysteresis loop for two values of the sweeping rate of the magnetic field in the $z$ direction. These results can be compared with the experimental results obtained in figure 3 in \cite{vergnani2012a}.

\subsection{A case of an arbitrary set of parameters with large fourth order term}
A system with arbitrary parameters is considered in order to investigate the effects of large values of fourth order terms. Here we consider the parameters: $S=5$, $D/k_B=-0.5$ K, $E/k_B=0$ K, $B_4^4/k_B=-1.5\times 10^{-2}$ K, $ B_x=0.001$ T, $B_y=B_4^0=B_4^2=B_4^3=0$.

The parameters in the transition rates $\gamma_{mm'}$ in (\ref{3.2}) used are: $3/(\pi \hbar^4 \rho c_s^5) = 1.831\times 10^3\;\mathrm{K}^{-5}\mathrm{s}^{-1}$, $D_1=D_2=0.5$ K, and $T=0.1$ K.

\begin{figure}[h!]
\begin{center}
\includegraphics[scale=0.65]{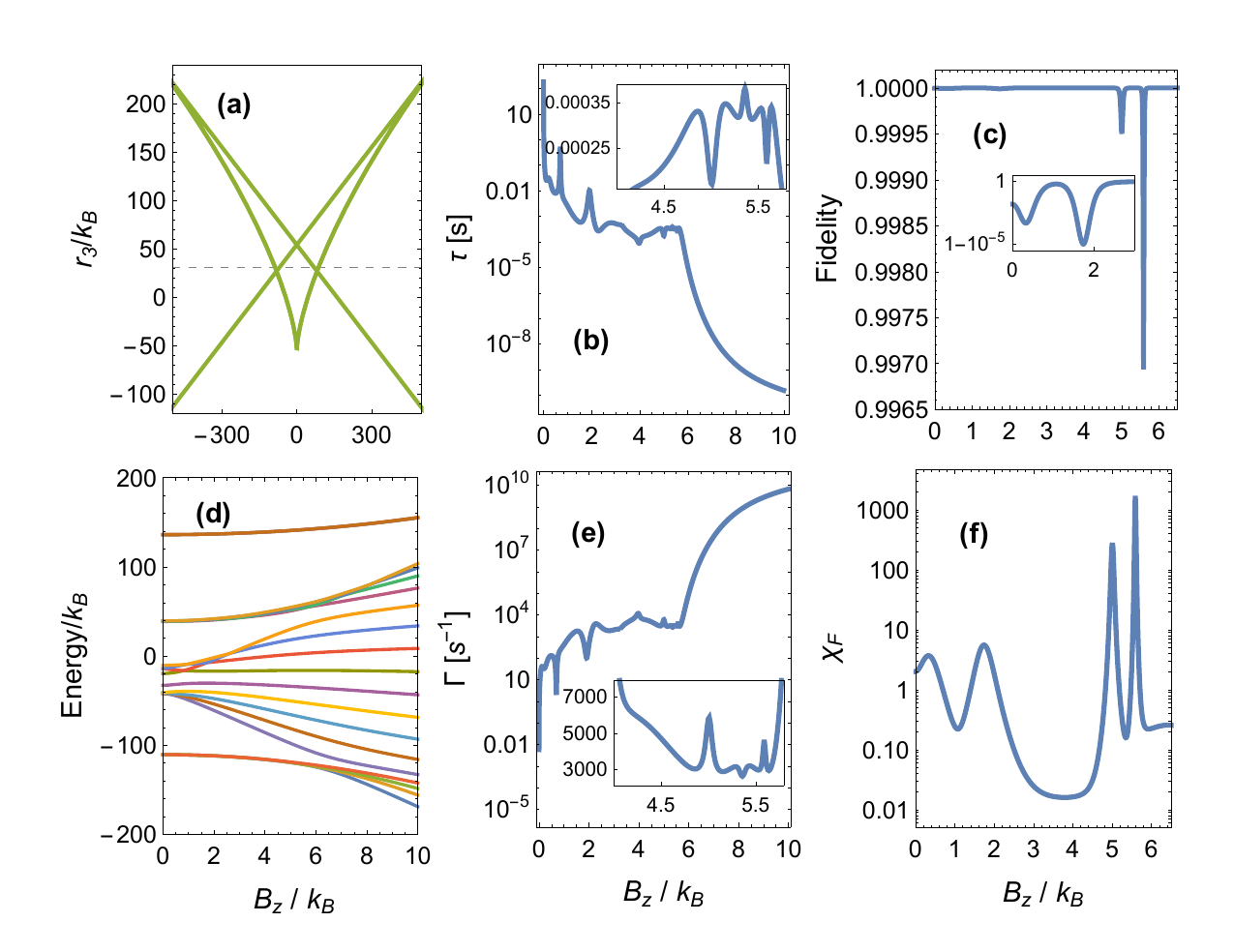}
\end{center}
\caption{(a) Bifurcation set in $(r_2/k_B,r_3/k_B)$ parameter space and the dotted line represents the parameters in the test case: $S=5$,  $D/k_B=-0.5$ K, $E/k_B=0$ K, $B_4^4/k_B=-1.5\times 10^{-2}$ K, $ B_x=0.001$ T, $B_y=B_4^0=B_4^2=B_4^3=0$; and $3/(\pi \hbar^4 \rho c_s^5) = 1.831\times 10^3\;\mathrm{K}^{-5}\mathrm{s}^{-1}$, $D_1=D_2=0.5$ K, and $T=0.1$ K.  (b) Relaxation time, (c) fidelity, (d) energy levels, (e) relaxation rate, (f) fidelity susceptibility as a functions of $B_z$. To calculate the fidelity we used $B_z=0.001$.}
\label{fig.3}
\end{figure}

In figure \ref{fig.3} we show the parameter space and we can see that in this case the horizontal dashed line corresponding to the parameter values crosses a butterfly catastrophe structure of the bifurcation set. The energy levels are shown in figure \ref{fig.3}(d) were the ground state energy has strong avoided level crossings for non-zero values of the magnetic field. This is reflected in the relaxation time and relaxation rate in figures \ref{fig.3}(b) and \ref{fig.3}(e), respectively, were in the insets we can see the respective drops and peaks about $B_z\approx 5$ T and $B_z\approx 5.6$ T. This structure can also be seen in the fidelity and fidelity susceptibility in figures \ref{fig.3}(c) and \ref{fig.3}(f), respectively, were the drops in the fidelity and the peaks of the fidelity susceptibility appear more clearly.

\section{Triple well potentials}\label{section6}
The spin Hamiltonian in consideration has two free parameters, namely the $x$ and $z$ components of magnetic field, while the parameters of {\it higher order terms are fixed} by the system of study. However it should be noted that the methods described above allow for the study of QPTs in parameter spaces involving free parameters of higher order terms in a similar fashion.

In this section we will perform a theoretical study and explore the complete parameter space for regions of interesting structural stability. Instead of freely varying the Hamiltonian parameters, by constructing first the semiclassical potential, and the separatrices in parameter space, we can have an idea of the behaviour and structure of the energy levels of the Hamiltonian, and it is this aspect where the usefulness and advantage of the methodology resides.

\begin{figure}[h!]
\begin{center}
\includegraphics[scale=0.56]{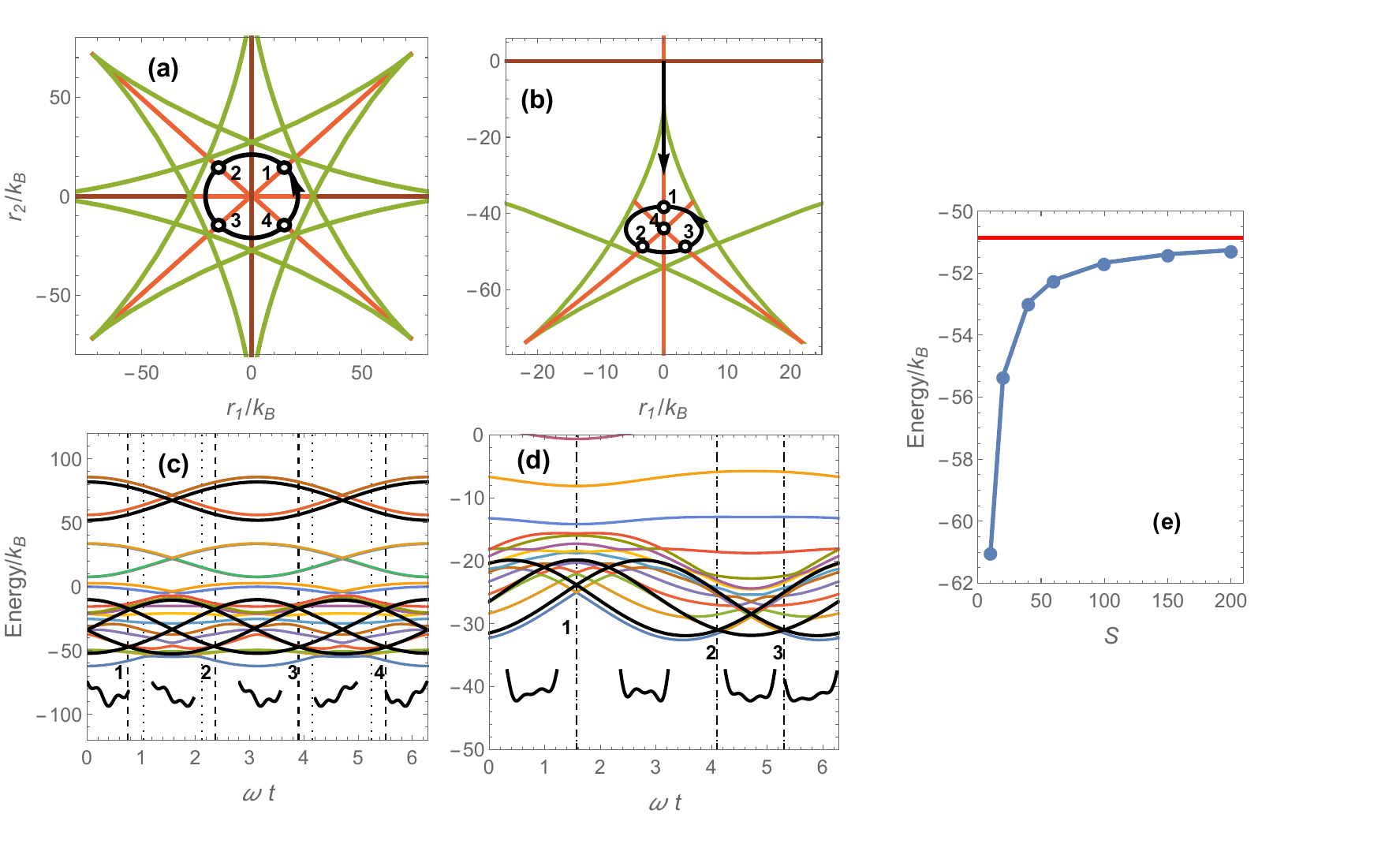}
\end{center}
\caption{(a) and (b) Parameter space $(r_1,r_2)$ for the cases I and II, respectively. The bifurcation set is depicted in green and Maxwell set in red (minima) and dark red (maxima). (a) The varying magnetic field $\vec{B}=(15.63\cos \omega t ,0,15.63\sin \omega t )$ is shown as a black circle with an arrow. The white dots where the Maxwell set for the global minimum is crossed are labelled ${\bd 1}$ to ${\bd 4}$. (b) The varying magnetic field $\vec{B}=(4.47\cos \omega t ,0,4.47\sin \omega t -32.9 )$ is shown as a black circle with an arrow. The white dots where the Maxwell set for the global minimum is crossed are labelled ${\bd 1}$ to ${\bd 3}$, and the white dot labelled ${\bd 4}$ correspond to the point where the semiclassical potential has three equally deep potential wells. (c) and (d) Energy levels as a function of $\omega t$, and in black lines the minima and maxima of the semiclassical potential are also plotted. In (c) and (d) the avoided level crossings of the ground state are indicated with a dotted vertical line, and the crossing with the Maxwell set with a dashed vertical line. At the bottom, schematic pictures of the semiclassical potential for each region are shown. (e) Ground state energy as a function of $S$ for the parameter values used in (a) for $\omega t=0.4$. The red horizontal line represents the semiclassical minimum.}
\label{fig.4}
\end{figure}

Two example cases are presented next where triple well potentials are found:

\begin{itemize}

\item[Case I:] Parameters: $S=10$, $D=-0.5$ K, $E=0.04$ K, $B_4^4=-0.007$ K, $B_4^0=B_4^2=B_4^3=0$. In figure \ref{fig.4}(a) we show the parameter space $(r_1,r_2)$ with the bifurcation set in green and the Maxwell set in red (minima) and dark red (maxima). This particular shape of the bifurcation set results from the overlap of two butterfly catastrophes one for each of the critical points $\phi_c=0,\pi$. This overlap is a consequence of the large value of the fourth order parameters $B_4^4$, which result in a large value of $r_4$.

\item[Case II:] Parameters: $S=10$, $D=0.4$ K, $E=-0.03$ K, $B_4^0=-0.00012$ K, $B_4^2=B_4^3=B_4^4=0$. In this case the parameter space consists of two cusps along the $r_1$-axis and two butterfly catastrophes along the $r_2$-axis. In figure \ref{fig.4}(b) we show one of the butterfly separatrix in parameter space $(r_1,r_2)$. At the intersection of two Maxwell set lines we find that the semiclassical potential has three equally deep potential wells.

\end{itemize}

\subsection{Magnetic field trajectories}
The magnetic field components $x$ and $z$ are free and we can use them to create a trajectory
in the parameter space, i.e., we can manipulate the system such that it passes from one phase
to another one. We will do it by a time varying magnetic field.

A quantum phase transition occurs when the parameters $r_1$ and $r_2$ are varied from one region to another, crossing a Maxwell set for the global minimum. We consider the variation of the magnetic field to have the following form: $\vec{B}=(|B|\cos \omega t + B_{x,0},0,|B|\sin \omega t + B_{z,0})$. In figures \ref{fig.4}(a) and \ref{fig.4}(b) this can be seen as the black circles, with the arrow indicating its direction: For case I we used $\vec{B}=(15.63\cos \omega t ,0,15.63\sin \omega t )$, and for case II we used $\vec{B}=(4.47\cos \omega t ,0,4.47\sin \omega t -32.9)$. The white dots indicate the points where a Maxwell set for the global minimum is crossed. The effect of crossing the Maxwell set can be seen in figures \ref{fig.4}(c) and \ref{fig.4}(d), where the energy levels as a function of $\omega t$ are plotted. Here we can see how an avoided level crossing of the ground state (dotted line) occurs near the crossing of the Maxwell set (dashed line). In black lines we plotted the maxima and minima of the semiclassical potential as a function of $\omega t$, and at the bottom an schematic picture of the semiclassical potential for each region is shown.

An additional trajectory is considered in figure \ref{fig.4}(a), where $B_x$ is fixed at zero or almost zero, and the component $B_z$ is varied passing through the point ${\bd 4}$, which correspond to the value of $B_z$ for which the semiclassical potential of the system has three equally deep stable wells.

In figure \ref{fig.4}(e) we have plotted the value of the ground state for one value of the magnetic field at $\omega t=0.4$ for case I as a function of $S$. As $S$ increases the ground state approaches the value of the global minimum of the semiclassical potential shown as an horizontal red line.

\begin{figure}[h!]
\begin{center}
\includegraphics[scale=0.8]{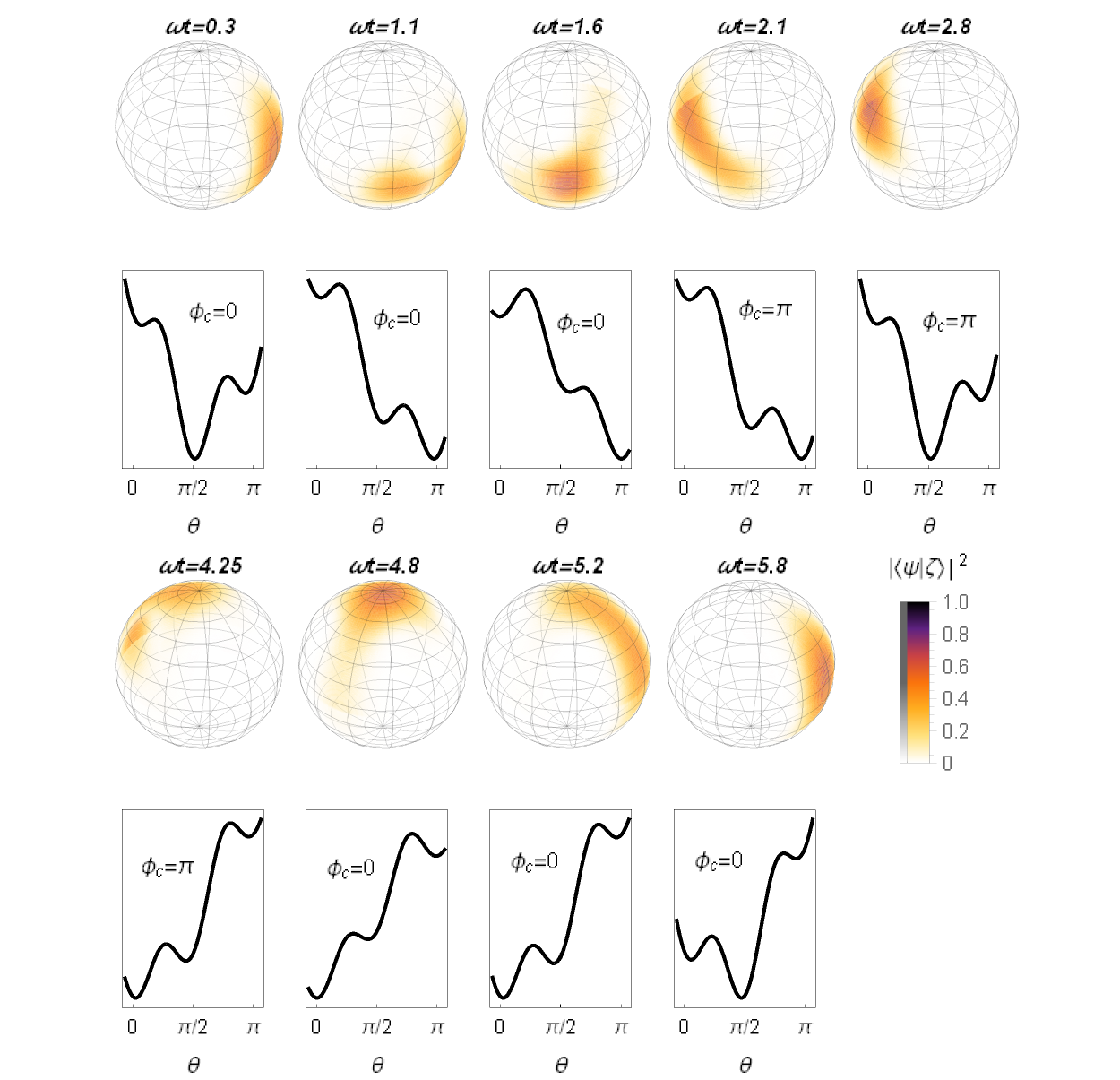}
\end{center}
\caption{Squared amplitude $|\langle\psi_1|\zeta \rangle |^2$ of the ground state eigenvector plotted on the Bloch sphere for various values of $\omega t$ for the trajectory shown in figure \ref{fig.4}(a). Below each sphere the corresponding semiclassical potential as a function of $\theta$ is shown.}
\label{fig.5}
\end{figure}

\subsection{Bloch sphere visualization}\label{section7}

As shown in the schematic representation of the semiclassical potential in each of the phases,as the Maxwell set is crossed, the deepest potential well changes from one to another. In this subsection we present a way to visualize the localization of the eigenstates of the Hamiltonian across these quantum phase transitions. Using the expression of the eigenvectors of the Hamiltonian in the basis of $|M\rangle$ in (\ref{1.3}) and the definition of the coherent states in (\ref{2.1}) we can define the complex function $\langle \psi_k | \zeta \rangle$ as the projection of the eigenvector on the coherent states:
\begin{equation}\label{7.1}
\fl \langle\psi_k|\zeta \rangle 
= \sum_{M=-S}^S  c_{k,M}^{*}\left(\frac{(2S)!}{(S+M)!(S-M)!}\right)^{1/2}\left(\cos \frac{\theta}{2}\right)^{S-M}\left(e^{-\mathrm{i} \phi}\sin \frac{\theta}{2}\right)^{S+M},
\end{equation}
where according to the definition of the coherent states $(\theta,\phi)$ are the angular coordinates on the Bloch sphere. The latitudes of the Bloch sphere are interpreted as the different spin projections $M$, with the south pole corresponding to $M=S$ and the north pole to $M=-S$, and for the case when $S$ is an integer we have that $M=0$ corresponds to the equator. The projection of the coherent states onto the eigenstates has been treated in \cite{lopezmoreno2014a}.

To show the information provided by (\ref{7.1}) we will apply it to the trajectory shown in figure \ref{fig.4}(a). In figure \ref{fig.5} we show $|\langle\psi_1|\zeta \rangle |^2$ for the ground state for various values of $\omega t$, to depict points before, during and after a transition. Note, that for some values of $\omega t$ the coherent state is localized to a definite area
on the Bloch sphere. Thus, controlling $\omega t$, stopping when a certain value is reached, we can
shift the coherent state at a certain area. In other words, we have a switch.

To complement this vision, in figure \ref{fig.5}, we also show the semiclassical potential as a function of $(\theta,\phi_c)$. At the start, $\omega t=0.3$, the ground state is localized at the equator, with $\theta \approx \pi /2$ and $\phi=0$, or in cartesian coordinates about $(x=1,y=0,z=0)$. As the magnetic field moves to $\omega t =1.1$ we can see the localization of the ground state starts to shift to the south pole. We can interpret this as a competition between the two wells close in energy, until, at $\omega t=1.6$ the well at about $\theta \approx \pi$ is the deepest. In turn, in figure \ref{fig.5} for that value, now the eigenstate of the ground state is localized in the south pole. As $\omega t$ increases the ground state continues to travel along the Bloch sphere: It started in the equator at about $(x=1,y=0,z=0)$, then it goes to the south pole, then to the equator at about $(x=-1,y=0,z=0)$, then to the north pole, and finally back to its starting point.

\begin{figure}[h!]
\begin{center}
\includegraphics[scale=0.48]{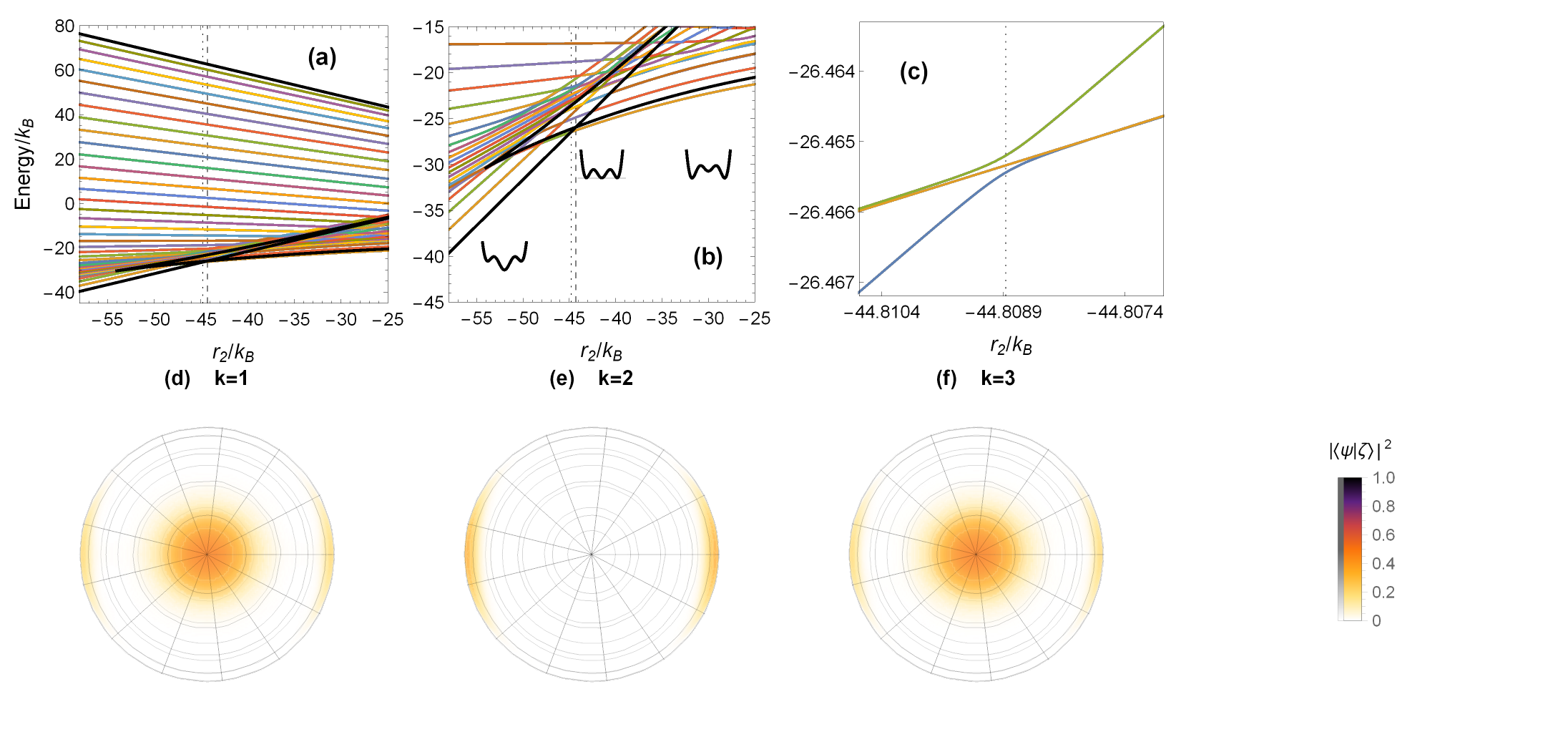}
\end{center}
\caption{(a), (b) and (c) Energy levels, for $S=20$, as a function of $r_2$ for the trajectory that crosses point ${\bd 4}$, which corresponds to the point where the potential has three equally deep wells, shown in figure \ref{fig.1}(b). The vertical dashed line corresponds to the semiclassical triple point at $r_2=-44.3$. The vertical dotted line corresponds to the avoided level crossing of the ground state at $r_2\approx -44.80887$. (d), (e) and (f): Squared amplitude $|\langle\psi_k|\zeta \rangle |^2$ plotted on the Bloch sphere, viewed from the top of the north pole, for the first three states at the avoided level crossing of the ground state.}
\label{fig.6}
\end{figure}

Another interesting feature of these example cases is the presence of the triple point ${\bd 4}$ in figure \ref{fig.4}(b), where the semiclassical potential has three equally deep wells. In figure \ref{fig.6} we show the structure of the eigenstates as a trajectory in parameters passes through that point. In the top row we plotted the energy levels as a function of $r_2$, and show a zoom view of the avoided level crossing at the triple point. At the bottom of the middle plot schematic pictures of the potential at the different regions are shown. The crossing of the semiclassical triple point is depicted as a dashed vertical line at $r_2=-44.3$, while the avoided level crossing of the ground state is depicted as a dotted vertical line at $r_2\approx -44.80887$. The minima and maxima of the semiclassical potential is plotted as a function of $r_2$ and shown as black lines. In the bottom row the $|\langle\psi_k|\zeta \rangle |^2$ functions are shown on the Bloch sphere, viewed from the top of the north pole, for the first three states. Here we can see how at the point of the avoided level crossing of the ground state, the ground state distribution has components at the north pole, and at opposite sides of the equator at $(x=1,y=0,z=0)$ and $(x=-1,y=0,z=0)$, consequence of the three wells. The second excited states also shares this behaviour, while the distribution of the first excited state is localized at both sides of the equator at $(x=1,y=0,z=0)$ and $(x=-1,y=0,z=0)$.

\section{Conclusions}
\label{conclusions}

In the present work we have discussed and studied many properties of spin Hamiltonians with terms up to fourth order and found how, by the variation of the external magnetic field, the system can go through the different phases in parameter space. We applied the powerful method of catastrophe theory.

To provide a real physical system, we studied two examples of single molecule magnets and showed that the sudden changes in fidelity and fidelity susceptibility are related with the inverse peaks in the relaxation rate and the peaks in the relaxation rate. For each of these examples the parameter space was constructed and it can be seen that the separatrices determine the structure of the eigenvalues of the Hamiltonian as a function of the magnetic field. We obtained a hysteresis for two magnetic molecules and demonstrated that the model applied is able to describe observation.

When considering large values of the fourth order parameters the separatrices in parameter space obtain a more complicated structure, which permits the inclusion of more phases, and in particular there exists regions where the semiclassical potential has three stability wells.

Next, we studied the manner on how to manipulate the position of the coherent state on the 
Bloch sphere, using a varying magnetic field. An oscillatory external magnetic field allows the system to travel through many of these different phases and the localization of the eigenstates can be visualized on the Bloch sphere, as well as how it changes along the sphere with the varying magnetic field, by performing a projection of the coherent states onto the eigenstates. This result can be viewed as the manipulation of a qubit state by means of a varying magnetic field, choosing the value of $\omega t$ in order to reach the desired orientation on the Bloch sphere. In one of the examples we found a point in parameter space which corresponds to the presence of three equally deep semiclassical potential wells. Using the visualization of the distribution  of the ground state on the Bloch sphere, we found that at the avoided level crossing point of the ground state with the first excited state, three separated regions on the sphere (north pole and $(\pm 1,0,0)$) have non zero contribution which is a consequence of the three competing stability points.

\ack
We acknowledge the financial support from PAPIIT-DGAPA IN117923, and IN116824.

\section*{References}

\providecommand{\newblock}{}

\end{document}